\documentclass{aa}

\input psfig.sty

\begin{document}
\newcommand{\mo}{$M_{\odot}$}
\newcommand{\mob}{$M_{\odot}\;$}

\title{Probability Distribution of Terrestrial Planets in Habitable Zones around Host Stars}

\author{Jianpo Guo\inst{1,2,3}
        \and Fenghui Zhang\inst{1,2}
        \and Xuefei Chen\inst{1,2}
        \and Zhanwen Han\inst{1,2}
        }

\offprints{Guo et al.}

\institute{National Astronomical Observatories/Yunnan Observatory,
Chinese Academy of Sciences,
           Kunming, 650011, P.R. China\\
           \email{guojianpo1982@hotmail.com}
          \and
          Key Laboratory for the Structure and Evolution of
Celestial Objects, Chinese Academy of Sciences,
           Kunming, 650011, P.R. China\\
           \and
           Graduate School of the Chinese Academy of Sciences,
Beijing, 100049, P.R. China\\
          }

\date{Published on Ap\&SS; Oct. 2009}

\abstract{With more and more exoplanets being detected, it is paid
closer attention to whether there are lives outside solar system. We
try to obtain habitable zones and the probability distribution of
terrestrial planets in habitable zones around host stars. Using
Eggleton's code, we calculate the evolution of stars with masses
less than 4.00 \mo. We also use the fitting formulae of stellar
luminosity and radius, the boundary flux of habitable zones, the
distribution of semimajor axis and mass of planets and the initial
mass function of stars. We obtain the luminosity and radius of stars
with masses from 0.08 to 4.00 \mo, and calculate the habitable zones
of host stars, affected by stellar effective temperature. We achieve
the probability distribution of terrestrial planets in habitable
zones around host stars. We also calculate that the number of
terrestrial planets in habitable zones of host stars is 45.5
billion, and the number of terrestrial planets in habitable zones
around K type stars is the most, in the Milky Way.

 }

\titlerunning{Terrestrial Planets in HZs}

\authorrunning{Guo et al.}

\maketitle

\section{Introduction}
Nowadays, more and more exoplanets have being indirectly detected,
and they were even directly detected last year (Marois et al.,
2008). And theories about planetary evolution point out that most of
stars own planets. It is paid closer attention to whether there are
lives around other stars, not only by astronomy filed, but also by
many other realms. A planet which is suited to life survive must
satisfy two factors, one is that it is a terrestrial planet, and the
other is that it must be in the habitable zone (HZ) of its host
star.

Terrestrial planets are rocky planets from one to ten Earth masses
with the same chemical and mineral composition as the Earth
(Valencia et al., 2006). However, others argued that the mass of
terrestrial planet could be down to about 0.3 Earth mass, to retain
its atmosphere over long geological timescales and to sustain
tectonic activity as required for the carbon-silicate cycle to
operate (Williams et al., 1997; Menou \& Tabachnik, 2003).

Typically, stellar HZ is defined as region near the host star, where
water at the surface of a terrestrial planet is in liquid phase
(eg., Hart, 1978; Kasting et al., 1993; Franck et al., 2000; Noble
et al., 2002; Jones et al., 2006). Previously, it is thought that
the boundary flux of HZ depends on luminosity (eg., Kasting et al.,
1993). However, it is pointed out that this flux also depends on
stellar effective temperature $T_{\rm eff}$ (Forget \&
Pierrehumbert, 1997; Williams \& Kasting, 1997; Mischna et al.,
2000; Jones 2004; Jones et al. 2006). As the greenhouse effect can
raise the mean temperature of terrestrial planets (Clube et al.,
1996), and the greenhouse effect is different for different
radiation. For example, the lower $T_{\rm eff}$, the more the
infrared fraction in luminosity, and the more this fraction, the
more the greenhouse effect for a given stellar flux (Jones et al.,
2006). Thus, the distances at both inner and outer HZ boundaries are
farther to host star, with lower $T_{\rm eff}$, than they would have
been if the $T_{\rm eff}$ effect is not taken into consideration.
And the distances are closer to host star with higher $T_{\rm eff}$.

It is well known that HZ widths for different stars are different.
For example, the HZ width for a M type star is only about one fifth
to one fiftieth of the HZ width for a G type star (Tarter et al.,
2007). However, it does not mean that the probability of terrestrial
planet in HZ around a M type star is also about one fifth to one
fiftieth of the probability around a G type star. Fortunately,
planetary evolution has been calculated, which matched well with the
observation data considering selective effects (Ida \& Lin, 2004,
2005; Schlaufman et al., 2009). Ida \& Lin (2005) gave the
distribution of semimajor axis and mass of planets predicted by the
Monte Carlo simulations. The masses of host stars are 0.20, 0.40,
0.60, 1.00 and 1.50 \mo, respectively. And this distribution can be
used to calculated the probability of terrestrial planets in HZs
around different host stars.

Jones et al. (2006) gave the HZ flux at both inner and outer
boundaries, as a function of $T_{\rm eff}$, but did not give the
flux for host stars. Using Eggleton's code, we calculate stellar
evolution and achieve the relationship between stellar mass and
$T_{\rm eff}$. We get the HZ flux, at both inner and outer
boundaries, for host stars at zero age main sequence (ZAMS) and at
the terminal of main sequence (TMS). And the HZ flux for host star
with mass 4.00 \mob is about five times more than that of host star
with mass 0.40 \mob at inner boundary and about four times at outer
boundary. Then we obtain the HZs of host stars with masses from 0.08
to 4.00 \mob at ZAMS and at TMS, taken the $T_{\rm eff}$ effect into
consideration.

Using the distribution of semimajor axis and mass of planets (Ida \&
Lin, 2005), we obtain the probability distribution of terrestrial
planets in HZs around host stars. Then, we calculate that the number
of terrestrial planets in habitable zones of host stars is 45.5
billion in the Milky Way. And the number of terrestrial planets in
HZs for M, K, G and F type stars are respectively as 11.548 billion,
12.930 billion, 7.622 billion and 5.556 billion, in the Milky Way.

The outline of the paper is as follows: we present some input
descriptions in Section 2, give our results in Section 3, present
some discussions in Section 4, and then finally in Section 5 we give
our conclusions.
\section{Input descriptions}
\subsection{The least lifetime of the main sequence for host stars}
Life emergence needs habitable environment with long enough time.
Jones et al. (2006) thought that it needed at least 1.0 Gyr. And
planets cover a presumed heavy bombardment phase as on Earth, at the
first 700 Myr (Jones, 2004; Lal, 2008). Hence, The lifetime of the
main sequence (MS) for host stars is at least 1.7 Gyr, according to
this issue.

However, this requirement may be too strict. The isotopic signature
of carbon granules in the Isua formation (more than 3.7 Gyr) is
interpreted as evidence for prokaryotic life (Rosing, 1999). And
there is evidence that there was liquid water on Earth about 4.3 Gyr
ago (Mojzsis et al., 2001). It is well known that the solar age is
about 4.57 Gyr (Bahcall et al., 1995). These mean that archaic life
had come forth when our solar was about 800 Myr old, and the
environment on Earth may be suited to biological evolution when
solar was about 200-300 Myr old. Hence, we adopt that the least
lifetime of MS for host stars is about 200 Myr.
\subsection{Input physics of stellar evolution}
We use the stellar evolution code of Eggleton (1971, 1972, 1973),
which has been updated with the latest input physics over the last
three decades (Han et al., 1994; Pols et al., 1995, 1998). We set
the convective overshooting parameter, $\delta_{\rm OV}=0.12$ (Pols
et al., 1997; Schr\"{o}der et al., 1997). We adopt metal mixture by
Grevesse \& Sauval (1998).

We use OPAL high temperatures opacity tables (Iglesias \& Rogers,
1996; Eldridge \& Tout, 2004) in the range of
$4.00<\mathrm{log(}\mathit{T}\mathrm{/K)}\leq 8.70$, and the new
Wichita state low temperature molecular opacity tables (Ferguson et
al., 2005) in the range of
$3.00\leq\mathrm{log(}\mathit{T}\mathrm{/K)}\leq 4.00$. And we have
made the opacity tables match well with Eggleton's code (Chen \&
Tout, 2007; Guo et al., 2008).

In our calculation, the value of metallicity is 0.02. As the
lifetime of MS for star with mass 4.00 \mob is about 183 Myr, we
calculate the evolution of stars with masses less than 4.00 \mo.
Stars with masses less than 0.50 \mob are spaced by
$\Delta{M}=0.05M_{\odot}$, and stars with masses more than 0.50 \mob
are spaced by $\Delta{M}=0.10M_{\odot}$.
\subsection{The initial mass function of stars}
In the Milky Way, it is well known that the number of stars with
different masses is different. In our calculation, we adopt the
initial mass function of Kroupa et al. (1993).
\begin{equation}
\xi(M)=\Bigl\{\matrix{c_{1}M^{-1.3}, & \ \ 0.08 \leq M < 0.5, \cr
c_{2}M^{-2.2}, & \ \ 0.5 \leq M < 1.0, \cr c_{2}M^{-2.7}, & \ \ 1.0
\leq M < \infty, \cr} \label{mdis}
\end{equation}
where $\xi(M)\mathrm{d}M$ is the probability of stars with masses
between $M$ and $M+\mathrm{d}M$, in solar units. The mass function
is normalized according to
\begin{equation}
\int_{0.08}^{\infty} \xi(M)\mathrm{d}M=1, \label{mdis}
\end{equation}
so that $\mathrm{c_{1}}=0.255508$ and $\mathrm{c_{2}}=0.138704$.
\section{Results}
\subsection{Fitting formulae for stellar luminosity and radius}
The fitting formulae of stellar luminosity and radius have been
completed by formers (Tout et al., 1996; Hurley et al., 2000).
However, the studies before 2005 used the old Wichita state low
temperature molecular opacity tables (Alexander \& Ferguson, 1994),
and there are some errors for the old opacity tables (Weiss et al.,
2006). Therefore, it is required to fit these formulae over again.
\subsubsection{Stellar luminosity and radius at ZAMS}
For stellar luminosity and radius at ZAMS, we fully use the former
shapes (Tout et al., 1996), and achieve a series of new fitting
coefficients.
\begin{eqnarray}
L_{\rm ZAMS}= \hspace{156pt}\nonumber \\
{\frac{a_{1}M^{5.5}+a_{2}M^{11}}
{a_{3}+M^{3}+a_{4}M^{5}+a_{5}M^{7}+a_{6}M^{8}+a_{7}M^{9.5}}},
\label{mdis}
\end{eqnarray}
\vspace{-5.0mm}
\begin{eqnarray}
R_{\rm ZAMS}= \hspace{179pt}\nonumber \\
{\frac{a_{8}M^{2.5}+a_{9}M^{6.5}+a_{10}M^{11}+a_{11}M^{19}+a_{12}M^{19.5}}
{a_{13}+a_{14}M^{2}+a_{15}M^{8.5}+M^{18.5}+a_{16}M^{19.5}}},
\label{mdis}
\end{eqnarray}
where $L_{\rm ZAMS}$ and $R_{\rm ZAMS}$ are stellar luminosity and
radius at ZAMS, respectively, both in solar units. And $M$ is
stellar mass, also in solar units, the same as Eqs. in what follows.
The fitting coefficients are seen in Tab. 1, the same as Eqs. (5),
(6) and (14). And stellar $T_{\rm eff}$ can be obtained from
$L=4{\pi}R^{2}{\sigma}T_{\rm eff}^{4}$. Hence, we can achieve the
relationship between stellar mass and $T_{\rm eff}$ at ZAMS, seen in
Fig. 1.
\begin{table}[tb]
 \caption{The fitting coefficients for Eqs. (3), (4), (5), (6) and (14),
             eg., $a_{20}=2.281557\times10^{2}$.}
 \label{table:1}
 \centering
 \begin{tabular}{ccc}
  \hline
  Names & \multicolumn{2}{c}{Values} \\
  \hline
  $a_{1}$-$a_{2}$  &$\hspace{7pt}5.799011(-1)$&$\hspace{7pt}1.975333(+1)$\\
  $a_{3}$-$a_{4}$  &$\hspace{7pt}4.062986(-4)$&$\hspace{7pt}9.883841(+0)$\\
  $a_{5}$-$a_{6}$  &$\hspace{7pt}1.737514(+1)$&$\hspace{7pt}2.540493(+0)$\\
  $a_{7}$-$a_{8}$  &$-1.117310(-1)$           &$\hspace{7pt}1.704116(+1)$\\
  $a_{9}$-$a_{10} $&$\hspace{7pt}6.263268(+1)$&$\hspace{7pt}8.677444(+1)$\\
  $a_{11}$-$a_{12}$&$-3.413124(+0)$           &$\hspace{7pt}5.031012(+0)$\\
  $a_{13}$-$a_{14}$&$\hspace{7pt}1.209713(-1)$&$\hspace{7pt}3.072570(+1)$\\
  $a_{15}$-$a_{16}$&$\hspace{7pt}1.553054(+2)$&$\hspace{7pt}8.999767(-1)$\\
  $a_{17}$-$a_{18}$&$\hspace{7pt}2.323133(+1)$&$-5.638277(+1)$\\
  $a_{19}$-$a_{20}$&$\hspace{7pt}5.744080(+1)$&$\hspace{7pt}2.281557(+2)$\\
  $a_{21}$-$a_{22}$&$\hspace{7pt}7.253992(-1)$&$\hspace{7pt}1.308659(+2)$\\
  $a_{23}$-$a_{24}$&$\hspace{7pt}7.658900(+0)$&$-2.603165(+3)$           \\
  $a_{25}$-$a_{26}$&$\hspace{7pt}1.567244(+1)$&$\hspace{7pt}1.617766(+0)$\\
  $a_{27}$-$a_{28}$&$\hspace{7pt}2.598481(+3)$&$-5.173614(-3)$           \\
  $a_{29}$-$a_{30}$&$\hspace{7pt}6.812527(+0)$&$-2.372715(+0)$           \\
  $a_{31}$-$a_{32}$&$-3.766615(+0)$           &$\hspace{7pt}2.528181(+0)$\\
  $a_{33}$-$a_{34}$&$\hspace{7pt}5.607562(+0)$&$-7.590386(-3)$           \\
  $a_{35}$-$a_{36}$&$-1.874321(-1)$           &$\hspace{7pt}3.915412(+0)$\\
  $a_{37}$-$a_{38}$&$\hspace{7pt}4.871765(+0)$&$\hspace{7pt}6.316666(-2)$\\
  $a_{39}$-$a_{40}$&$\hspace{7pt}7.910890(-1)$&$\hspace{7pt}5.999996(-1)$\\
  $a_{41}$         &$\hspace{7pt}2.052252(-1)$\\
  \hline
 \end{tabular}
\end{table}
\begin{figure}
\psfig{file=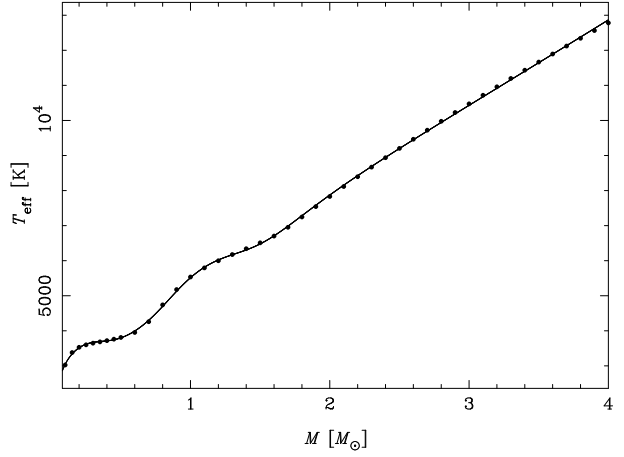,height=6.0cm,bbllx=586pt,bblly=28pt,bburx=31pt,bbury=770pt,clip=,angle=270}
\caption{The relationship between stellar mass and $T_{\rm eff}$ at
ZAMS. Points are stellar data calculated by Eggleton's code, and the
line is taken from fitting formula, the same as Fig. 2.}
\label{ised}
\end{figure}
\subsubsection{Stellar luminosity and radius at TMS}
For stellar luminosity and radius at TMS, we modify the former
shapes (Hurley et al., 2000) and fit again, and the expressions are
\begin{eqnarray}
L_{\rm TMS}= \hspace{151pt}\nonumber \\
{\frac{a_{17}M^{2.5}+a_{18}M^{3}+a_{19}M^{4}+a_{20}M^{a_{23}+1.8}}
{a_{21}+a_{22}M^{5}+M^{a_{23}}}}, \label{mdis}
\end{eqnarray}
\begin{equation}
R_{\rm
TMS}=\Bigl\{\matrix{\frac{r10+r11M^{r12}+r13M^{r14}}{r15+M^{r16}}, &
\ \ M < 1.4, \cr \cr
\frac{r17+r18M^{r19}+r20M^{r19+1.5}}{r21+M^{5}}, \ \ & \ \ \ M \geq
1.4, \cr} \label{mdis}
\end{equation}
where both $L_{\rm TMS}$ and $R_{\rm TMS}$ are in solar units.
\subsection{HZs of host stars}
The inner HZ boundary is determined by the loss of water via
photolysis and hydrogen escape. And the outer HZ boundary is
determined by the condensation of $\rm CO_2$ crystals out of the
atmosphere (von Bloh et al., 2007). Jones et al. (2006) gave the HZ
flux at both inner and outer boundaries, as a function of $T_{\rm
eff}$.
\begin{equation}
\frac{S_{\rm in}}{S_{\odot}}=4.190\times10^{-8}T_{\rm
eff}^{2}-2.139\times10^{-4}T_{\rm eff}+1.296, \label{mdis}
\end{equation}
\vspace{-5.0mm}
\begin{equation}
\frac{S_{\rm out}}{S_{\odot}}=6.190\times10^{-9}T_{\rm
eff}^{2}-1.319\times10^{-5}T_{\rm eff}+0.2341, \label{mdis}
\end{equation}
where $S_{\odot}$ is solar constant and $T_{\rm eff}$ is in Kelvin.
As we have obtained the relationship between stellar mass and
$T_{\rm eff}$, we can obtain the HZ boundary flux of host stars with
masses from 0.08 to 4.00 \mo, seen in Fig. 2. The more massive is
the host star, the higher is $T_{\rm eff}$, and the more is the HZ
flux at both inner and outer boundaries.

As $L=4{\pi}d^{2}S$, and the three parameters are just as
$L_{\odot}$, AU and $S_{\odot}$, for our Earth. Hence, the distances
at both inner and outer HZ boundaries are given by
\begin{equation}
\frac{d_{\rm in}}{AU}=\bigg[\frac{L/L_{\odot}}{S_{\rm
in}/S_{\odot}}\bigg]^{1/2}, \label{mdis}
\end{equation}
\begin{equation}
\frac{d_{\rm out}}{AU}=\bigg[\frac{L/L_{\odot}}{S_{\rm
out}/S_{\odot}}\bigg]^{1/2}. \label{mdis}
\end{equation}
\begin{figure}
\psfig{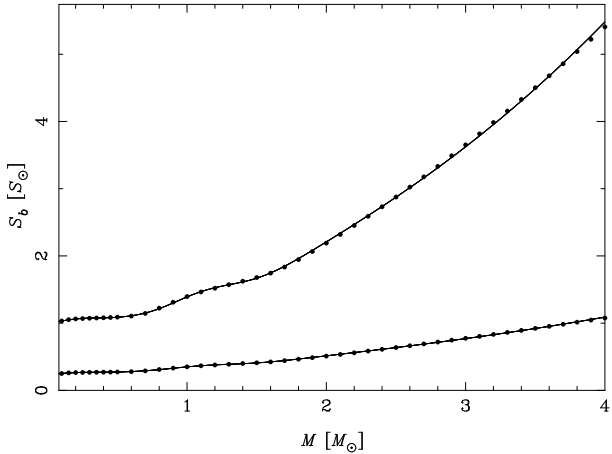}
\caption{The HZ flux at both inner boundary (upper line) and outer
boundary (lower line), considering the $T_{\rm eff}$ effect.}
\label{ised}
\end{figure}
According to the data of Fig. 2 and Eqs. (3), (9) and (10), we can
obtain the distances of HZs around host stars with masses from 0.08
to 4.00 \mob at ZAMS, seen in Fig. 3. Using the same method, we
achieve the distances of HZs around host stars at TMS, also seen in
Fig. 3.
\begin{figure}
\psfig{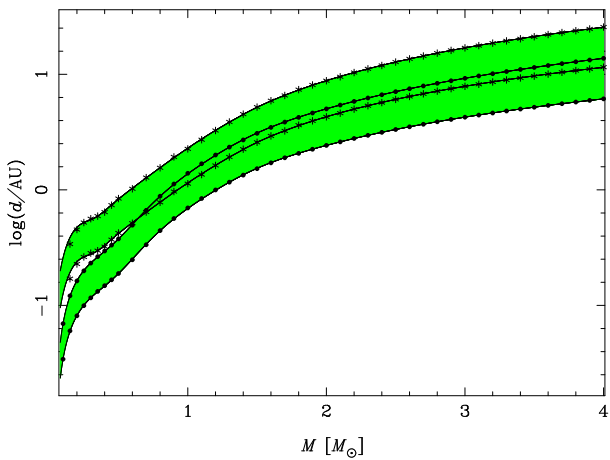}
\caption{The HZs of host stars at ZAMS and at TMS, taken the $T_{\rm
eff}$ effect into consideration. The lines with points denote the HZ
boundary distances at ZAMS, and the lines with asterisks denote the
HZ boundary distances at TMS. Points and asterisks are stellar data
taken from Eggleton's code, and the lines are taken from fitting
formulae.} \label{ised}
\end{figure}
\subsection{Probability distribution of terrestrial planets in HZs around host stars}
We take that terrestrial planet is the planet with mass from 0.3 to
10 Earth masses (Williams et al., 1997; Menou \& Tabachnik, 2003;
Valencia et al., 2006), and semimajor axis is the distance between
the planet and its host star. Using the data of Fig. 2a in Ida \&
Lin (2005), we count the number of terrestrial planets whose
semimajor axis lie in the HZs of the host stars at ZAMS, noted as
$np(M)$. The masses of host stars are 0.20, 0.40, 0.60, 1.00 and
1.50 \mo, respectively. And we denote the probability of terrestrial
planets in HZs around host stars as $p(M)$. As the evolution of a
total of 20\,000 planets is calculated for each run, the
relationship between $p(M)$ and $np(M)$ is
\begin{equation}
p(M)=\mathrm{kp}\times np(M)/20000. \label{mdis}
\end{equation}

In order to obtaining the value of $\mathrm{kp}$, we need the data
about planets in our solar systems. The number of planets with
masses more than 0.1 Earth mass, for the host star with mass 1.00
\mo, is 6429, according to the data of Fig. 2a in Ida \& Lin (2005).
And there are 8 planets with masses more than 0.1 Earth mass, in our
solar system. Hence, we can get the value of $\mathrm{kp}$ from
\begin{equation}
8=\mathrm{kp}\times 6429/20000, \label{mdis}
\end{equation}
so that $\mathrm{kp}=24.88723$. Therefore, we get the linear
relationship between $p(M)$ and $np(M)$, just as
\begin{eqnarray}
p(M)=24.88723\times np(M)/20000\hspace{11pt}\nonumber \\
=1.2443615\times10^{-3}\times np(M). \label{mdis}
\end{eqnarray}

For the host star with mass 0.20 \mo, the distances at both inner
and outer HZ boundaries are 0.081288 and 0.162912 AU, respectively.
And the $np(M)$ is 52, so that $p(M)=0.064707$, according to Eq.
(13). Using the same method, we also get the $np(M)$ and the $p(M)$
for host stars with masses 0.40, 0.60, 1.00 and 1.50 \mo, seen in
Tab. 2. According to the $p(M)$ for these five stars, we obtain
$p(M)$ around host stars with masses from 0.08 to 4.00 \mo, seen in
Eq. (14) and Fig. 4.
\begin{table}[tb]
\caption []{The distances at both inner and outer HZ boundaries, and
the values of $np(M)$ and $p(M)$, for host stars with masses 0.20,
0.40, 0.60, 1.00 and 1.50 \mo} \label{table:2} \centering
\begin{tabular}{ccccc}
\hline Stars & $d_{\mathrm{in}}$ & $d_{\mathrm{out}}$ & $np(M)$ &
$p(M)$\\
\hline
0.20 &0.081288 &0.162912  &52   &0.064707 \\
0.40 &0.148050 &0.295742  &87   &0.108259 \\
0.60 &0.248416 &0.494705  &312  &0.388241 \\
1.00 &0.697398 &1.391183  &745  &0.927049 \\
1.50 &1.525654 &3.085378  &794  &0.988023 \\
\hline
\end{tabular}
\end{table}
\begin{equation}
p(M)=\Bigl\{\matrix{a_{36}M^{a_{37}}+a_{38}, & \ \ M \leq 0.60, \cr
a_{39}(M-a_{40})^{0.1}+a_{41}, \ \ & \ \ \ M > 0.60. \cr}
\label{mdis}
\end{equation}
\begin{figure}
\psfig{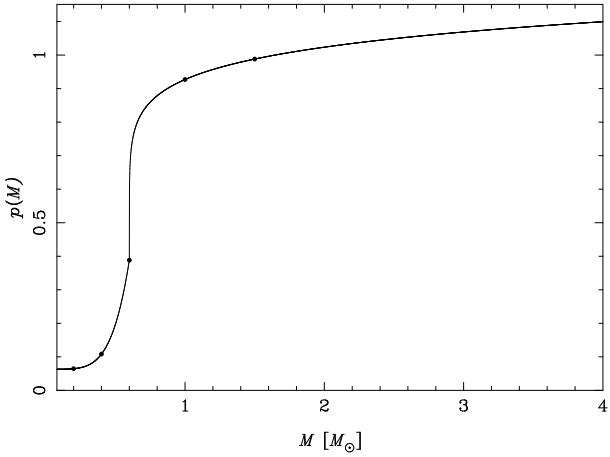}
\caption{The probability distribution of terrestrial planets in HZs
around host stars. The data of five points is obtained from the
distribution of semimajor axis and mass of planets (Ida \& Lin 2005)
and our results about HZs of host stars at ZAMS. The data of line is
fitted from the five points.} \label{ised}
\end{figure}

It may be too rough that $p(M)$ is obtained by extrapolation, with
stellar mass more than 1.50 \mo. However, it is seen that $p(M)$
increases with increasing stellar mass and the increasing trend
becomes slower and slower, for host stars with masses from 0.60 to
1.50 \mo. And $p(M)$ lasts this increasing trend, for host stars
with masses more than 1.50 \mo. In addition, the value of $\xi(M)$
for stars with masses more than 1.50 \mob in the Milky Way is very
low. Therefore, it is acceptable that the formula is fitted from
five points, although it is not very perfect.
\subsection{Probability distribution of terrestrial planets in HZs around host stars in the Milky Way}
We use $Prob(M)$ to denote the probability of terrestrial planets in
HZs around host stars in the Milky Way. Therefore, $Prob(M)$ is
equal to $p(M)$ multiplied by $\xi(M)$ seen in Eq. (1).
\begin{equation}
Prob(M)=p(M)\times\xi(M), \label{mdis}
\end{equation}
also intuitively seen in Fig. 5. And $Prob(M)\mathrm{d}M$ is the
probability of terrestrial planets in HZs around all the host stars
with masses between $M$ and $M+\mathrm{d}M$ in the Milky Way.
\begin{figure}
\psfig{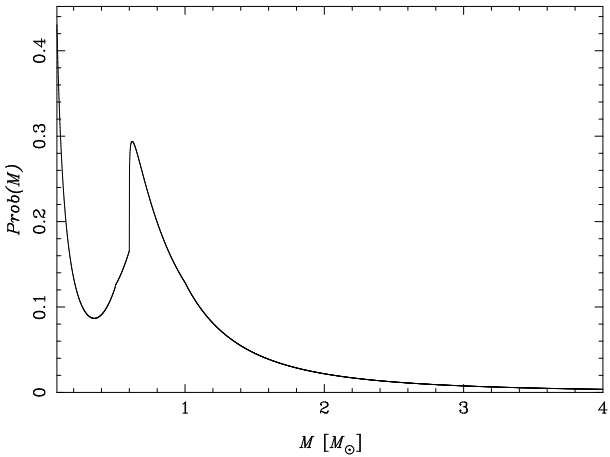}
\caption{The probability distribution of terrestrial planets in HZs
around host stars in the Milky Way.} \label{ised}
\end{figure}

When stellar mass is less than 0.35 \mo, $p(M)$ increases gently,
but $\xi(M)$ decreases dramatically. Hence, $Prob(M)$ decreases with
increasing stellar mass. When stellar mass is from 0.35 to 0.62 \mo,
$p(M)$ increasing rate is more than $\xi(M)$ decreasing rate, so
that $Prob(M)$ increases with increasing stellar mass. And $Prob(M)$
meets a peak value, with stellar mass 0.62 \mo. When stellar mass is
more than 0.62 \mo, $\xi(M)$ decreasing rate is more than $p(M)$
increasing rate, so that $Prob(M)$ decreases with increasing stellar
mass.

\subsection{The number of terrestrial planets in the Milky Way}
The number of stars in the Milky Way is about 300 billion. And
observations tell us that upwards of 50\% of the stars populating
galaxies are expected to be in binary or higher-order multiple
systems (eg., Duquennoy \& Mayor, 1991; Richichi et al., 1994).
Previously, it was generally thought that these systems are not
suited to life survive. Recently, it was pointed out that
terrestrial planets may be stable in binary systems for a long time
(eg., David et al., 2003; Fatuzzo et al., 2006; Saleh \& Rasio,
2009). Hence, the number of host stars that may be suitable for life
survive in the Milky Way is about 200 billion, noted $N$ as the
follows. Therefore, the number of terrestrial planets in HZs of host
stars in the Milky Way is
\begin{equation}
N_{\rm planet}=N\int_{0.08}^{4.00}
p(M)\xi(M)\mathrm{d}M=4.5507\times10^{10}. \label{mdis}
\end{equation}

As the mass range of M, K, G and F type stars are about from 0.08 to
0.50 \mo, from 0.50 to 0.80 \mo, from 0.80 to 1.05 \mob and from
1.05 to 1.40 \mob (Grie$\beta$meier et al., 2005), respectively.
Therefore, the number of terrestrial planets in HZs for M, K, G and
F type stars are 11.548 billion, 12.930 billion, 7.622 billion and
5.556 billion, respectively.

These mean that the number of terrestrial planets around K type
stars is the most in the Milky Way. Because both the number of K
type stars and the $p(M)$ around K type stars are great. And the
$p(M)$ around M type stars is very low, although the number of M
type stars is huge. On the contrary, the number of G type stars is
minor, although the $p(M)$ around G type stars is high.

In order to estimating the number of terrestrial planets with lives
or even intelligent lives in the Milky Way, we use the critical step
model of biological evolution (Cater \& McCrea, 1983; Watson, 2008).
The model assumes that biological evolution needs sequentially pass
several critical steps. And the successful probability for each step
is very low, eg., 10\% (Watson, 2008). The five step model is
perfect, as it matches well with the observed date. Watson (2008)
pointed out that the five critical steps are life emergence (primary
life), prokaryotes to eukaryotes, asexual clones to sexual
populations, cell differentiation (complex life) and primate
societies to human societies (intelligent life).

As the number of terrestrial planets in HZs of host stars in the
Milky Way is about 45.5 billion, and biological evolution needs
enough time. Therefore, the number of terrestrial planets with
primary lives in the Milky Way is about 4.3 billion. And the number
of terrestrial planets with complex lives and intelligent lives in
the Milky Way are about 3.7 million and 0.36 million, respectively.
\section{Discussion}
There are many planets with semimajor axis about 0.03 AU around host
stars with different masses, most of which are terrestrial planets
(Ida \& Lin, 2005). And the HZs of host stars with masses about 0.10
\mob is approximately just as 0.03 AU. Hence, there must be many
terrestrial planets in the HZs for these stars. Considering that
there is huge number of stars with masses about 0.10 \mo, the
probability of terrestrial planets in HZs around these stars must be
very high in the Milky Way.

In the Monte Carlo simulations of planet formation processes before
2008 (eg., Ida \& Lin, 2005), it neglected the effects of type I
migration of protoplanetary embryos due to their tidal interaction
with their nascent disks. And it is found that the type I migration
provides a self-clearing mechanism for planetesimals in the
terrestrial planet region (Ida \& Lin, 2008), which may modulate the
distribution of semimajor axis and mass of planets.

Ultraviolet radiation is a double-edged sword to life. If it is too
strong, the terrestrial biological systems will be damaged. And if
it is too weak, the synthesis of many biochemical compounds can not
go along (Buccino et al., 2006). In our future work, we are also
plan to study the ultraviolet habitable zones of host stars.
\section{Conclusion}
Firstly, we obtain the HZs of host stars with masses form 0.08 to
4.00 \mob at ZAMS and at TMS, considering the $T_{\rm eff}$ effect.
Secondly, we give the probability distribution of terrestrial
planets in HZs of host stars. Thirdly, we calculate the number of
terrestrial planets in HZs of host stars is 45.5 billion in the
Milky Way, and find that the number of terrestrial planets around K
type stars is the most in the Milky Way. Finally, we present
discussions about the host stars with masses about 0.10 \mob and the
effects of type I migration, and introduce our future work. One may
also send any special request to \it{guojianpo1982@hotmail.com}
\normalfont{or} \it{guojianpo16@163.com}\normalfont{.}

\begin{acknowledgements}
This work is supported by the National Natural Science Foundation of
China (Grant Nos. 10773026, 10821061 and 2007CB815406), the Chinese
Academy of Sciences (Grant No. KJCX2-YW-T24) and Yunnan Natural
Science Foundation (Grant No. 06GJ061001). Jianpo Guo thanks Prof
Ida, Prof Watson and Prof Jones for help.
\end{acknowledgements}

{}
\end{document}